\documentclass[aps,twocolumn,showpacs,superscriptaddress,floatfix]{revtex4}
\usepackage{graphicx}

\begin{document}

\title{The theory of frequency shifts as a function of the spin location,
using the oscillating cantilever-driven adiabatic reversals technique}
\author{G.P. Berman}
\affiliation{Theoretical Division and CNLS, MS B213, 
Los Alamos National Laboratory, Los Alamos, NM 87545} 

\author{F. Borgonovi}
\affiliation{Dipartimento di Matematica e Fisica, Universit\'a Cattolica, 
via Musei 41, 25121 Brescia, Italy}
\affiliation{INFN, Sezione di Pavia, Italy}

\author{V.I. Tsifrinovich}
\affiliation{IDS Department, Polytechnic University, Six Metrotech
Center, Brooklyn, New York 11201}

\date{\today} 

\begin{abstract} 

{The theory of the oscillating cantilever-driven adiabatic reversals
(OSCAR)
in magnetic resonance force microscopy
(MRFM) is extended to describe 
the  relation between an  external magnetic field and a dipole 
magnetic field  for an arbitrary location of the single spin. An analytical
estimate for the OSCAR MRFM frequency shift is derived and 
shown to be in  excellent agreement with numerical 
simulations. The dependence of the
frequency shift on the position of the spin relative to the cantilever
has characteristic maxima and minima which can be used to determine
the  spin location experimentally.
}

\end{abstract} 


\maketitle 

\section{Introduction}

The oscillating cantilever driven adiabatic reversals (OSCAR) technique
in magnetic resonance force microscopy  (MRFM) introduced in \cite{stipe}
has been used to successfully detect  a single electron 
spin below the surface of a solid \cite{rug1}. In the OSCAR MRFM technique
the vibrations of the cantilever tip (CT) with an attached ferromagnetic 
particle in presence of a {\it rf} magnetic field cause the periodic reversals
of the effective magnetic field acting on the single electron spin.
If the conditions of adiabatic motion are satisfied the spin follows the 
effective magnetic field. The back action of the spin on the CT causes a
small frequency shift of the CT vibrations, which can be measured with high 
precision.

The quasiclassical theory of OSCAR MRFM has been developed in \cite{bkt}.
This theory contains two important limitations. First, it assumes
that the external magnetic field $\vec{B}_{ext}$ at the spin is much greater than the
dipole field $\vec{B}_d$ produced by the ferromagnetic particle. 
In real experiments, in order to increase the frequency shift $\delta\omega_c$,
one has to decrease the distance between the CT and the spin to values
where the dipole field becomes sometimes greater than the external field\cite{rug1}.
Second, it was assumed in \cite{bkt} that the spin is located in the
plane of the cantilever vibrations. Thus, the quasiclassical theory should
be extended in order to describe both an arbitrary relation between 
$\vec{B}_{ext}$ and $\vec{B}_d$ and an arbitrary location of the spin. 
This extension is presented in our paper.

A single spin is a quantum object which  must be described 
using  quantum theory. The quantum theory of OSCAR MRFM has 
been developed in \cite{bbt} with the same limitations as the quasiclassical 
theory. It was found, as may be expected, that the frequency shift 
$\delta\omega_c$ in quantum theory may assume only two values 
$\pm |\delta\omega_c|$ corresponding to the two directions of the spin 
relative to the effective magnetic field. The value of $|\delta\omega_c|$
in quantum theory is the same as the maximum frequency shift calculated
using  quasiclassical theory (where it can take any value between 
$-|\delta\omega_c|$ and $|\delta\omega_c|$). Thus, to calculate the
quantum frequency  shift, it    is reasonable to use quasiclassical instead of
quantum theory.

\section{Equations of motion}

\begin{figure}
\includegraphics[scale=0.4]{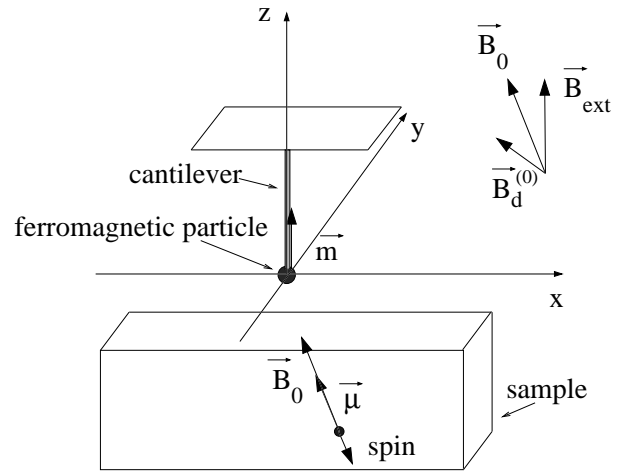}
\caption{
MRFM setup. The equilibrium position of the spin
and the cantilever with a spherical ferromagnetic particle.
$\vec{m}$ is the magnetic moment of the ferromagnetic particle,
$\vec{\mu}$ is the magnetic moment of the spin, $\vec{B}_{ext}$,
$\vec{B}_d^{(0)}$ and $\vec{B}_0$ are 
respectively the external permanent
magnetic field, the dipole field on the spin,
and the net magnetic field. In general the vectors 
$\vec{B}_d^{(0)}$ and $\vec{B}_0$ do not lie in the $x-z$ plane.
}
\label{exp}
\end{figure}

We consider the MRFM setup shown in Fig.~\ref{exp}.
The CT oscillates in the $x-z$ plane. The origin is placed
at the equilibrium position of the center
of the ferromagnetic particle).
Note that here we ignore the static displacement of the CT caused by the 
magnetic force of the spin.
The magnetic moment of the spin, $\vec{\mu}$  shown in Fig.~\ref{exp},
points initially in the direction of the magnetic field $\vec{B}_0$,
which corresponds to the equilibrium position of CT. (See Eq.~(\ref{bfie})). 
We assume now that the {\it rf} magnetic field $2 \vec{B}_1$ is linearly 
polarized in the plane which is perpendicular to $\vec{B}_0$.
(Later we will generalize this to an arbitrary direction of polarization).
The dipole magnetic field $\vec{B}_d$ is given by:

\begin{equation}
\vec{B}_d = \frac{\mu_0}{4\pi} \frac{3(\vec{m}\cdot\vec{n})\vec{n}-\vec{m}}
{r_v^3},
\label{dipf}
\end{equation}
where $\vec{m}$ is the magnetic moment of the ferromagnetic particle
pointing in the positive $z$-direction, $r_v$ is the (variable) distance
between the moving CT and the stationary spin and $\vec{n}$ is the unit vector
pointing from the CT to the spin. We define:

\begin{equation}
r_v = \sqrt{ (x-x_c)^2 +y^2+z^2},
\label{co1}
\end{equation}
\begin{equation}
\vec{n} = \left( \frac{x-x_c}{r_v}, \
 \frac{y}{r_v}, \  \frac{z}{r_v}\right),
\label{coo}
\end{equation}
where $x,y,z$ are the spin coordinates, and  $x_c$ is the CT-coordinate
(i.e. the coordinate of the center of the ferromagnetic particle).
At the equilibrium, the net magnetic field at  the spin is,

\begin{equation}
\vec{B}_0 = \vec{B}_{ext} + \vec{B}_d^{(0)},
\label{bfie}
\end{equation}
\begin{equation}
\vec{B}_d^{(0)} = \frac{ 3 m \mu_0 }{4\pi  r^5}
\left( zx, \ zy, \ z^2-
\frac{r^2}{3} \right),
\label{bfi1}
\end{equation}
\begin{equation}
\vec{B}_{ext} = (0 , \ 0, \ B_{ext}),
\label{bfi2}
\end{equation}
where $r=\sqrt{x^2+y^2+z^2}$.
In the approximation which is linear in  $x_c$, the magnetic field $\vec{B}_d$
changes by the value of $\vec{B}_{d}^{(1)}$:

\begin{equation}
\vec{B}_d^{(1)} = - (G_x, G_y, G_z)  \ x_c,
\label{bp1}
\end{equation}
\begin{equation}
(G_x,G_y,G_z)  = \frac{3 m \mu_0}{4\pi r^7} \left( z(r^2-5x^2), 
-5xyz,  x(r^2- 5z^2) \right),
\label{bp2}
\end{equation}
where $(G_x,G_y,G_z)$ describes the gradient of the magnetic field at 
the spin location at $x_c=0$:

\begin{equation}
(G_x,   G_y,  G_z) = 
\left( \frac{\partial B_{d}^x}{\partial x},
\frac{\partial B_{d}^y}{\partial x},  \frac{\partial B_{d}^z}{\partial x}
\right).
\label{gg}
\end{equation}

(Note that the magnetic field and its gradient depend 
on the CT coordinate $x_c$).
Next we consider the equation of motion for the spin magnetic
moment $\vec{\mu}$ in the system of coordinates rotating
with the {\it rf} field at frequency $\omega$ about the magnetic field
$\vec{B}_0$. (The $\tilde{z}$ axis of this new system points in the direction
of $\vec{B}_0$). We have:

\begin{equation}
\begin{array}{lll}
\vec{\dot{\mu}} &= - \gamma \vec{\mu} \times \vec{B}_{eff},\\
&\\
\vec{B}_{eff} &= \left( B_1,\   0,\  B_0 -(\omega/\gamma) -
x_c \sum_i G_i \cos\alpha_i\right),\\
&\\
\cos\alpha_i &= B_0^i/B_0.\\
\end{array}
\label{beff}
\end{equation}
Here $\alpha_i$, ($i=x,y,z$) 
 are the angles between the direction of the magnetic
field $\vec{B}_0$ and the axes $x,y,z$ of the laboratory system of
coordinates; and  $\gamma$ is the gyromagnetic ratio of the electron spin.
 ($\gamma$ is the absolute value of the gyromagnetic ratio).
We ignore the transverse components of the dipole field 
$\vec{B}_d$ because they represent the fast oscillating terms in the
rotating system of coordinate. Also we consider only the rotating 
component of the {\it rf} magnetic field.

The equations of motion for  the CT  can be written,

\begin{equation}
\ddot{x}_c + \omega^2_c x_c = F_x/m^*,
\label{eom}
\end{equation}
where $\omega_c$ and $m^*$ are the frequency and the effective mass of the 
CT and  $F_x$ is the magnetic force acting on the ferromagnetic particle
on CT. We consider the CT oscillations in the laboratory system of 
coordinates. Ignoring fast oscillating terms in the laboratory system, we
obtain:
\begin{equation}
F_x = -\mu_{\tilde{z}} \sum_i G_i \cos\alpha_i.
\label{fdx}
\end{equation}

Next, we will use the following units: for time $ 1/\omega_c$, for magnetic 
moment $\mu_B$, for magnetic field $\omega_c/\gamma$, for length
the characteristic distance $L_0$ between CT and the spin, for force
$k_cL_0$, where $k_c= m^* \omega_c^2$ is the effective CT spring constant.
Using these units, we derive the following dimensionless equations of motion:

\begin{equation}
\displaystyle
\begin{array}{lll}
&\dot{\vec{\mu}} = -  \vec{\mu} \times \vec{B}_{eff},\\
&\\
&\ddot{x}_c +  x_c = F_x,\\
&\\
&\vec{B}_{eff} = \left( B_1,  0, \Delta-\beta {\cal G} x_c\right),\\
&\\
&F_x = -\alpha \beta {\cal G} \mu_{\tilde{z}}, \\
&\\
&\Delta = B_0 -\omega,\\
&\\
&\displaystyle {\cal G} = \frac{1}{r^7}[ z(r^2-5x^2)\cos\alpha_x
-5xyz\cos\alpha_y
\\
&\\
&+x(r^2-
5z^2) \cos\alpha_z].\\
\end{array}
\label{adi}
\end{equation}
The parameters $\alpha$ and $\beta$ are given by:

\begin{equation}
\alpha =\frac{\mu_B \omega_c}{\gamma k_c L_0^2}, \qquad 
\beta = \frac{3\gamma\mu_0 m }{4\pi\omega_c L_0^3}.
\label{ab}
\end{equation}
Note that all quantities in Eq.~(\ref{adi}) are dimensionless, i.e. $x$ means 
$x/L_0$,  $\mu$ means $\mu/\mu_B$,  $B_0$ means $\gamma B_0/\omega_c$ 
and so on. 
In terms of dimensional quantities the parameter $\beta$ is the ratio of
the dipole frequency $\gamma B_d^{(0)}$ to the CT frequency
$\omega_c$, and the product $\alpha\beta$ is the ratio
of the static CT displacement $F_x/k_c$ to the CT-spin distance $L_0$.
The derived equations are convenient for both numerical simulations and
analytical estimates.

\begin{figure}
\includegraphics[scale=0.33]{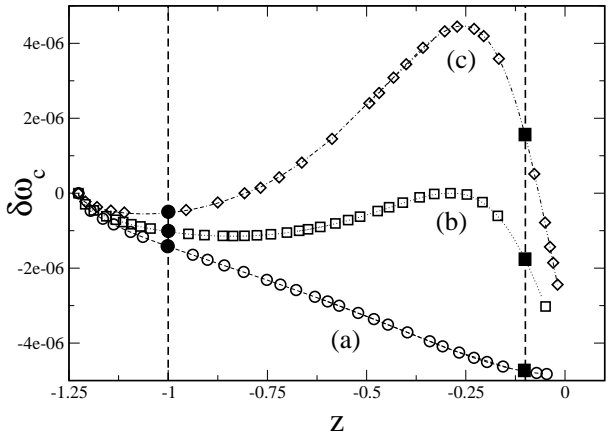}
\caption{
The OSCAR MRFM frequency  shift $\delta\omega_c(z)$ at the central
resonant surface  ($\Delta = 0$), for $x>0$. The symbols  show
the numerical data, the lines correspond to the estimate (\ref{dom})
for  (a) $y=0$ (circles), (b) $y=x/2$ (squares) and (c) $y=x$
(diamonds). Solid squares and circles  indicate frequency shifts
at the spin locations indicated in Fig.~\ref{cro}.
In all Figures the coordinates $x,y$ and $z$ are in units of $L_0$
and the frequencies are in units of $\omega_c$.
}
\label{osci}
\end{figure}

\section{The OSCAR MRFM frequency shift}

In this section we present the  analytical estimates and the numerical
simulations for the OSCAR MRFM frequency shift. When the CT oscillates,
the resonant condition $\omega = \gamma |\vec{B}_{ext} + \vec{B}_d|$
can be satisfied only if the spin is located inside the resonant 
slice which is defined by its boundaries:

\begin{equation}
|\vec{B}_{ext} + \vec{B}_d (x_c=\pm A)| = \omega/\gamma,
\label{res}
\end{equation}
where $A$ is the amplitude of the CT vibrations.
For an analytical estimate,  we assume that the spin is located 
at the central surface of
the resonant slice. In this case in Eq.~(\ref{adi}) $\Delta=0$.

To obtain an analytical estimate for the OSCAR MRFM
frequency shift we will assume
an ideal adiabatic motion and put $\vec{\dot{\mu}}=0$ in Eq.~(\ref{adi}).
Let the CT begin  its motion (at $t=0$) from the right  end 
position $x_c(0)=A$. Then the initial direction (i.e. at $t=0$) of the effective
magnetic field $\vec{B}_{eff}$ relative to the magnetic field
$\vec{B}_{ext}+\vec{B}_d$ and of the magnetic moment $\vec{\mu}$
depends on the sign of ${\cal G}$: $\vec{B}_{eff}$ and $\vec{\mu}$
have the same direction for ${\cal G} < 0$ and opposite directions 
for ${\cal G}>0$. Substituting the derived expression for 
$\mu_{\tilde{z}} \simeq - B^z_{eff}{\cal G}/|\vec{B_{eff}}||{\cal G}|$ 
into $F_x$ we obtain the following equation for $x_c$:

\begin{equation}
\ddot{x}_c +x_c \left\{ 1 + \frac{\alpha \beta^2 {\cal G}| {\cal G}| } 
{\sqrt{B_1^2 + (\beta {\cal G} x_c)^2 }} \right\}=0.
\label{simp}
\end{equation}
We solve this equation as in \cite{bkt}, using the
perturbation theory of Bogoliubov and Mitropolsky\cite{bogo}, and we find
the dimensionless frequency shift (see Appendix):

\begin{eqnarray}
\nonumber &&\delta\omega_c \simeq
\frac{2}{\pi} \frac{\alpha \beta^2 {\cal G} |{\cal G}|}
                    {\sqrt{B_1^2 +(\beta  {\cal G} A)^2 }}
\left\{ 1+\right. \\ 
&& \nonumber \\
&& \left. \frac{1}{2} \frac{B_1^2}{B_1^2+(\beta{\cal G} A)^2}
\Big[
\ln \Big( \frac{4\sqrt{B_1^2 + (\beta{\cal G} A)^2}}{B_1}\Big)  
+ \frac{1}{2} \Big]
\right\}.
\label{dom}
\end{eqnarray}
In typical experimental conditions we have $$B_1 \ll \beta {\cal G} A, $$ and 
Eq.~(\ref{dom}) transforms to the  simple expression

\begin{equation}
\delta\omega_c = \frac{2}{\pi} \frac{\alpha \beta {\cal G}}{ A}.
\label{dom1}
\end{equation}
One can see that the frequency shift is determined by the ratio of the static
CT displacement $F_x/k_c$ to the amplitude of the CT vibrations A.
We will also present Eq.~(\ref{dom}) in terms of dimensionless
quantities:

\begin{equation}
\frac{\delta\omega_c}{\omega_c} =
\frac{2\mu_B G_0}{ \pi A k_c},
\label{domdim}
\end{equation}
where
\begin{equation}
G_0=\sum_i G_i \cos\alpha_i. 
\label{gg0}
\end{equation}
Eqs.~(\ref{dom}) and (\ref{domdim}) 
represent an extension of the estimate derived in
\cite{bkt}. These equations are valid for any point on the central resonant 
surface and for any relation between $\vec{B}_{ext}$ and $\vec{B}_d$. 
It follows  from Eq.~(\ref{dom}) that $\delta\omega_c$ is an even 
function of $y$ and an odd function of $x$.

\begin{figure}
\includegraphics[scale=0.33]{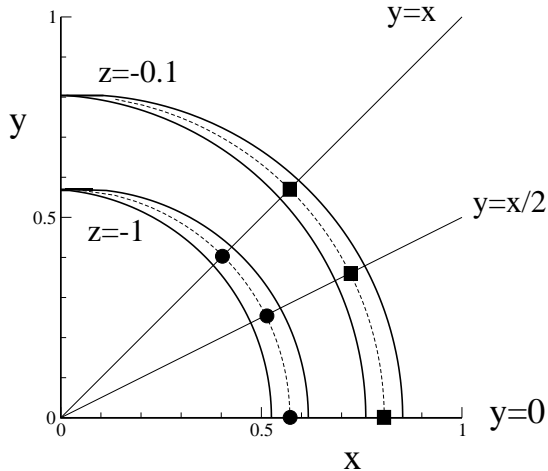}
\caption{Cross-sections of the resonant slice
for $z=-0.1$ and $z=-1$. The dashed lines show the intersection
between the cross sections and the central resonant surface.
The solid squares and circles  indicate spin locations
which correspond to the frequency shifts 
given by the same symbols  in Fig.~\ref{osci}.
}
\label{cro}
\end{figure}

In our computer simulations we have used the following parameters taken from 
experiments  \cite{rug1}: 
$$
\omega/2\pi = 5.5 \ kHz, \quad k_c= 110 \ \mu N/m, \quad A=16 \ nm,
$$
$$
B_{ext} = 30 \ mT, \quad \omega/2\pi = 2.96 \ GHz, \quad \omega/\gamma= 
106 \ mT,
$$
$$
|G_z| = 2\times 10^5 \ T/m, \quad B_1 = 300 \ \mu T, 
\quad L_0 \approx 350 \ nm.
$$

The corresponding dimensionless parameters are the following:
$$
\alpha = 1.35\times 10^{-13}, \quad \beta  = 1.07 \times 10^6, \quad 
A=4.6\times 10^{-2},
$$
$$
B_1 = 1.5 \times 10^3, \qquad B_{ext} = 1.53 \times 10^5, \qquad \omega= 
5.4\times 10^5.
$$

As  initial conditions we take:
$$
\vec{\mu}(0) = (0, 0, 1), \qquad 
x_c(0)  = A, \qquad \dot{x}_c (0) = 0.
$$

Below we describe the results of our computer simulations.
Fig.~\ref{osci} shows the frequency shift $\delta\omega_c$
as a function of the spin $z$-coordinate at the central resonant
surface ($\Delta = 0$). First, one can see an excellent agreement
between the numerical data and the analytical estimate 
(\ref{dom}). Second, as  expected, the maximum magnitude
of the frequency shift $|\delta\omega_c|$ can be achieved when
the spin is located in the plane of the CT vibrations $y=0$.
However, for $y=x$, it has almost the same magnitude $|\delta\omega_c|$
(with the opposite sign of $\delta\omega_c$).
Moreover, for $y=x$ the dependence 
$\delta\omega_c(z)$ has an extremum, which can be used 
for the measurement of the spin $z$-coordinate.
If the distance between the CT and the surface of the sample
can be controlled, then the ``depth'' of the spin location below 
the sample surface can be determined. (In all Figures, the coordinates
$x,y$ and $z$ are given in units of $L_0$, and the frequency shift
is in units of $\omega_c$.)

Fig.~\ref{cro}  shows the  cross-sections of the resonant
slice for $z=-0.1$ and $z=-1$.
The greater the distance from the CT, the smaller   
the cross-sectional area. The  solid squares and circles  in Fig.~\ref{cro}
show the spin locations which correspond to the 
frequency shifts indicated by the same symbols  in Fig.~\ref{osci}.

\begin{figure}
\includegraphics[scale=0.33]{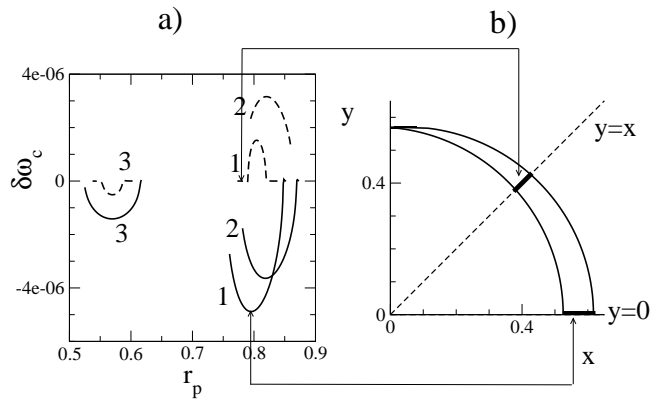}
\caption{a) The OSCAR MRFM frequency  shift $\delta\omega_c(r_p)$ inside
the cross-sectional area of the resonant slice
for $x>0$. The solid lines correspond to $y=0$ and the dashed lines
correspond to $y=x$. Lines are $1 - $, $z=-0.1$,
$2 - $, $z=-0.43$, and $3 - $, $z=-1$.
$r_p = (x^2+y^2)^{1/2}$.
b) the cross-section
of the resonant slice $z=-0.1$.
The bold segments show the spin locations which correspond to the lines
$1$  on $a)$.
}
\label{nf4}
\end{figure}

Fig.~\ref{nf4} demonstrates the ``radial'' dependence of the 
frequency shift $\delta\omega_c (r_p)$, where $r_p = (x^2+y^2)^{1/2}$.
The value of $r_p$ can be changed by the lateral displacement of
the cantilever. As one may expect, the maximum value of 
$|\delta\omega_c|$ corresponds to the central resonant surface.
The maximum becomes sharper as z decreases. Thus, a  small
distance between the CT and the sample surface
is preferable for the measurement of the radial position
of the spin. 

\begin{figure}
\includegraphics[scale=0.33]{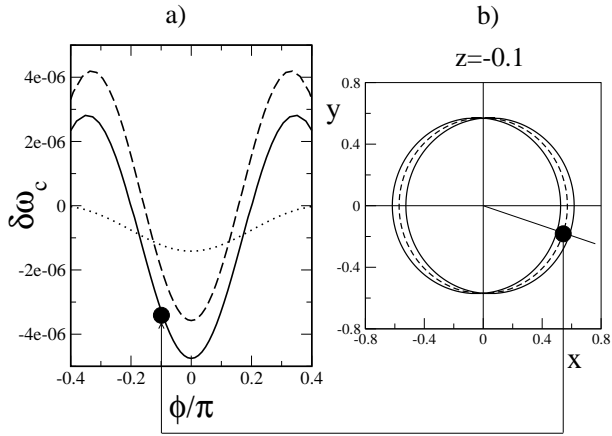}
\caption{a)  $\delta\omega_c (\phi)$, with
$\phi= \tan^{-1} (y/x)$ for the central resonant surface
and  $z=-0.1$ (full line);  $z=-0.43$ (dashed line), $z=-1$
(dotted line).
b) solid line shows the cross-section of the
resonant slice for $z=-0.1$. 
Dashed line shows the intersection between 
the plane $z=-0.1$ and 
the central resonant surface. The solid circle in b) shows 
the spin location $\phi/\pi=-0.1$ whose corresponding frequency shift
is marked by a solid circle   on a).
}
\label{dphi}
\end{figure}

Fig,~\ref{dphi} shows the ``azimuthal dependence'' of the
frequency shift $\delta\omega_c (\phi)$, where 
$\phi = \tan^{-1} (y/x)$ and the spin is located on the central
resonant surface. Note that for the given values
of $z$ and $\phi$, the coordinates $x$ and $y$ of the spin 
are fixed if the spin is located on the central resonant surface.
The value of $\phi$ can be changed by rotating the cantilever
about its axis. One can see the sharp extrema of the function 
$\delta\omega_c (\phi)$. Again, the small distance between the
CT and the sample is preferable for the measurement of the 
``azimuthal position'' of the spin.

\begin{figure}
\includegraphics[scale=0.33]{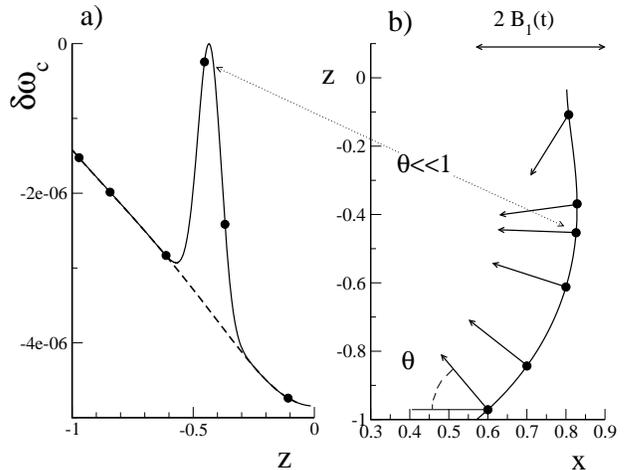}
\caption{ a) 
$\delta\omega_c (z)$ when the {\it rf} field $\vec{B}_1$
is parallel to the $x$-axis. The spin is located 
at the central resonant surface $y=0$, $x>0$. Solid line 
are numerical data, dashed line is the analytical estimate (\ref{dom}),
which assumes adiabatic motion of the spin 
magnetic moment
$\vec{\mu}$  parallel to 
$\vec{B}_{eff}$. For a few numerical points indicated as solid cirlces in a)
the corresponding $\vec{B}_0$ field is shown in   b).
b) solid line : intersection between the central resonant surface and the 
$x-z$ plane. Arrows show the magnetic field $\vec{B}_0$ on this intersection
at the  points indicated as solid circles in a).
The absolute value of the frequency shift $|\delta\omega_c|$ drops at the 
spin locations where $\vec{B}_0$ is approximately parallel to 
$\vec{B}_1$ $(\theta \ll 1)$. 
}
\label{pola}
\end{figure}

Finally, we consider the realistic case in which the direction of polarization
of the {\it rf} field $2\vec{B}_1$ is fixed in the laboratory system
of coordinates.
Now the angle $\theta$ between the direction of polarization
of $2\vec{B}_1$ and the field $\vec{B}_0$ depends 
on the spin coordinate because the magnitude and the direction of the 
dipole field $\vec{B}_d^{(0)}$ depend on the spin location.
To describe this case we ignore the component of $2\vec{B}_1$
which is parallel to $\vec{B}_0$, and change $B_1$ to $B_1 \sin\theta$
in all our formulas. As an example, Fig.~\ref{pola}
demonstrates 
 the dependence $\delta\omega_c(z)$
for the case in which the {\it rf} field is polarized
along the $x$-axis. One can see that in a narrow region of $z$
the  magnitude  of the
frequency shift sharply drops. This occurs because in this region the
magnetic field $\vec{B}_0$ is almost parallel to the $x$-axis.
Thus, the effective field $B_1 \sin\theta$ is small:
the condition of the adiabatic motion
$ \gamma [ B_1 \sin \theta ]^2 \gg | d\vec{B}_{eff}/dt |$
is not satisfied; and the spin
does not follow the effective magnetic field.
The dashed lines in Fig.~\ref{pola} correspond to the analytical estimate
(\ref{dom}) with the substitution $B_1 \to B_1 \sin\theta$: the analytical
estimate assumes adiabatic conditions, which are violated for small $\theta$.

The sharp drop of $|\delta\omega_c |$ could be observed either by the
change of the distance between the CT and the sample surface or by the
change of the direction of polarization of the {\it rf} field.

In any case this effect provides an  independent measurement
of the spin ``depth'' below the sample surface.

\section{Conclusion}

We have derived the quasiclassical equations of motion
describing the OSCAR technique in MRFM for arbitrary
relation between the external and dipole
magnetic fields and arbitrary location of a single spin. We have
obtained an analytical estimate of the OSCAR MRFM
frequency shift $\delta\omega_c$
which is in excellent agreement with numerical simulations. We have shown
that the dependence $\delta\omega_c$ on the position of spin relative to
the cantilever contains characteristic maxima and minima which can be used 
to determine the position of the spin. We believe that moving cantilever
in three dimensions, rotating it (or the sample)
about  the cantilever's  axis and changing the
direction of the polarization of the {\it rf} magnetic field, experimentalist
eventually will enable the determination of  the position of a single spin.
We hope that our work will help to achieve this goal.

\section*{Acknowledgments}

This work was supported by the Department of Energy (DOE) under
Contract No. W-7405-ENG-36, by the Defense Advanced Research Projects Agency 
(DARPA), by the National Security Agency (NSA),
and by the Advanced Research and Development Activity (ARDA).

\section{Appendix}

Eq.~(\ref{simp}) can be written in the following form:
\begin{equation}
\frac{d^2 x_c}{d\tau^2} + x_c = \epsilon f(x_c),
\label{aa1}
\end{equation}
where $\tau=\omega_c t$ is the dimensionless time,
\begin{equation}
f(x_c) =  \frac{\beta {\cal G} x_c}{\sqrt{B_1^2+(\beta {\cal G})^2 x_c^2}},
\label{aa2}
\end{equation}
and $\epsilon= - \alpha\beta|{\cal G}| $.

The  approximate solution of (\ref{aa1}), can be
written as \cite{bogo}: 

\begin{equation}
x_c(\tau)=a(\tau) \cos\psi(\tau)+ O(\epsilon),
\label{aa3}
\end{equation}
where in the first order in $\epsilon$,
$a(\tau)$ and $\psi(\tau)$ satisfy the equations:
\begin{equation}
\begin{array}{lll}
\displaystyle \frac{da}{d\tau} &= \epsilon P_1 (a)
+O(\epsilon), \\
&\\
\displaystyle \frac{d\psi}{d\tau} &= 1 + \epsilon Q_1 (a)
+O(\epsilon), \\
\end{array}
\label{aa4}
\end{equation}
and the functions $P_1(a)$ and $Q_1(a)$ are given by:

\begin{equation}
P_1(a) = -\frac{1}{2\pi} \int_0^{2\pi} f(a\cos\psi) \sin\psi \ d\psi,
\label{a5a}
\end{equation}
\begin{equation}
Q_1(a) = -\frac{1}{2\pi a} \int_0^{2\pi} f(a\cos\psi) \cos\psi \ d\psi. 
\label{a5b}
\end{equation}
On inserting the explicit expression (\ref{aa2})
for $f(a\cos\psi)$ one gets:
\begin{equation}
\label{a6a} 
P_1(a) = 0,
\end{equation}
\begin{equation}
\label{a6b} 
Q_1(a) = -\frac{2\beta {\cal G} }{\pi\sqrt{B_1^2 +(\beta {\cal G} a)^2}} 
\int_0^{\pi/2} \frac{(1-\sin^2 \psi)}{\sqrt{1-k^2\sin^2\psi}}
\ d\psi,
\end{equation}
where 
\begin{equation}
k^2 = \frac{(\beta {\cal G} a)^2}{ B_1^2 +(\beta {\cal G} a)^2}.
\label{aa7}
\end{equation}
Eq.~(\ref{a6b}) can be written as:
\begin{equation}
Q_1(a) = -\frac{2\beta {\cal G} }{\pi k^2\sqrt{B_1^2+(\beta {\cal G} a)^2}}
[ (k^2 -1) K(k) + E(k)],
\label{aa8}
\end{equation}
where $K(k)$  and $E(k)$
are the complete elliptic integrals of the first and second kind.
When $k \simeq  1$  elliptic
integrals can be approximated by:
\begin{equation}
K(k) \approx \ln \frac{4}{\sqrt{1-k^2}}
+\frac{1}{4}\left( \ln \frac{4}{\sqrt{1-k^2}} -\frac{1}{2}\right)
(1-k^2)  ,
\label{a9a}
\end{equation}
\begin{equation}
E(k) \approx 1 + \frac{1}{2}\left( \ln \frac{4}
{\sqrt{1-k^2}} -\frac{1}{2} \right) (1-k^2).
\label{a9b} 
\end{equation}
In the first order approximation the frequency shift is: 
\begin{eqnarray}
\nonumber && \delta\omega_c \simeq \epsilon Q_1(a) =
\frac{2}{\pi} \frac{\alpha \beta^2 {\cal G} |{\cal G}|}
                    {\sqrt{B_1^2 +(\beta  {\cal G} a)^2 }}
\left\{ 1+ \right. \\ 
\nonumber && \\
&& \left. \frac{1}{2}\frac{B_1^2}{B_1^2+(\beta{\cal G} a)^2}
\Big[
\ln \Big( \frac{4\sqrt{B_1^2 + (\beta{\cal G} a)^2}}{B_1}\Big)  
+\frac{1}{2}
\Big]
\right\}
.
\label{aa9}
\end{eqnarray}
In the  approximation $a \approx A$, one obtains Eq.~(\ref{dom}).

\end{document}